\documentclass[aps,prd,nofootinbib,twocolumn,longbibliography]{revtex4-1}
\usepackage[english]{babel}
\usepackage[autostyle, english = british]{csquotes}
\usepackage[CJKbookmarks, pdftex, bookmarksnumbered, bookmarksopen, colorlinks, citecolor=blue, linkcolor=blue]{hyperref}
\usepackage{amsmath,amssymb}
\usepackage{graphicx}
\usepackage{enumitem}
\setlist[enumerate]{topsep=0pt,parsep=-1mm,leftmargin=5mm,}
\def\be{\begin{equation}}
\def\ee{\end{equation}}

\begin{document}

\title{\large How causation is rooted into thermodynamics}

\author{Carlo Rovelli}
\affiliation{Aix-Marseille University, Universit\'e de Toulon, CPT-CNRS, F-13288 Marseille, France.}\affiliation{Department of Philosophy and the Rotman Institute of Philosophy, 1151 Richmond St.~N London  N6A5B7, Canada}\affiliation{Perimeter Institute, 31 Caroline Street N, Waterloo ON, N2L2Y5, Canada}

\begin{abstract} 
\noindent The notions of \emph{cause} and \emph{effect} are widely employed in science. I discuss why and how they are rooted into thermodynamics. The entropy gradient (i) explains in which sense interventions affect the future rather than the past, and (ii) underpins the time orientation of the subject of knowledge as a physical system.   Via these two distinct paths, it is this gradient, and only this gradient, the source of the time orientation of causation, namely the fact the cause comes \emph{before} its effects. 
\end{abstract}

\maketitle

\section{Introduction}

\noindent Common causation is asymmetric: the eruption of the Vesuvio was the cause of the destruction of Pompei, not the other way around. And is time oriented: the cause happens \emph{before} its effect.  There is a curious tension (but no real contradiction) between the fact that such time-oriented notion of causation is ubiquitous in our sciences, and yet we know that the elementary equations describing nature do not distinguish the past from the future.   I clarify this tension here.  

The solution of the apparent tension is that causation is rooted into thermodynamics \cite{Lewis1979,Dowe1975}.  The arrow of causation (from cause to effect) is a consequence of the thermodynamic arrow (from lower to higher entropy). Since the first points to the (the temporal direction we call) future, so does the second. The relation between the two arrows, however, is subtle and multi-faced. It is both related to the simple existence of thermodynamically  irreversible phenomena, and to the time orientation of the agent employing the notion of causation, and is intertwined with the epistemic and the agential temporal arrows (we know the past better than the future; we can act on the future but not on the past).  Here I disentangle the relations  between these time orientations, and show that they are all dependent on the thermodynamic one.  

Many of the ideas in this note have been anticipated in \cite{Ismael2022a}.  For related ideas, see \cite{Carroll}

\section{What is a cause?}

The notion of `cause' is utilized extensively in science as in everyday life.  But making precise sense of what is meant by causation is notoriously tricky.  `Cause' can have different meanings. Already in antiquity, Aristotle famously  distinguished four kinds of causes (material, formal, efficient and final \cite{Falcon2011}) and Buddhist philosophy six \cite{Dhammajoti2009}.   Here I focus on the most common notion (Aristotle's efficient causation). Examples are: the eruption of the Vesuvio has been the cause of the destruction of Pompei, my push is the cause of the door opening, shots by Lee Oswald caused the death of John Kennedy, smoking is a cause of cancer, a stone falling into a pond causes waves, and so on.   I do not start with a sharp definition, because clarifying what is the implicitly definition utilised when we talk about these causes is one of the ingredients of what follows. 

Among scientists working in fundamental theoretical physics, it is commonly assumed that causation (in the sense of these examples) does not play any role in the elementary physical description of the world.\footnote{By ``causation" I mean here what is called ``strong causation" in \cite{Adlam2022}.  This is the oriented notion of causation of our intuition, where $A$ causes $B$ and not viceversa. This should not be confused with relativistic causality, namely the fact that that there cannot be space-like influences. The first is asymmetric, the second symmetric. In quantum field theory, the second is expressed by the fact that operators supported in space-like separated regions commute.  To be ``causally connected" in this symmetrical sense  is a much weaker notion.  And so is the more general symmetric notion of causal connection considered in \cite{Adlam2022}, analysing non standard quantum causal models. To be causally connected (in this weaker sense, or in the sense of relativistic causality) is a condition for causation to be possible. Strong causation is much more than this, as clarified below, and is directional.}  In fact, no fundamental elementary law describing the physical world that we have found is expressed in terms of causes and effects. Rather, laws are expressed as regularities, in particular describing correlations, among the natural phenomena. Furthermore, these correlations do not distinguish past from future: they do not have any orientation in time.\footnote{There is a common mistake in the literature: stating that parity-charge violation and CPT invariance pick a preferred time direction. This is wrong. What to call positive or negative charge and left or right are conventions, hence nothing measurable picks a preferred one among two histories related by CPT. This is the same mistake as the claim that the Maxwell equations pick a direction of time because a time reversed solution is  a solution only by flipping the sign of the electric field. It stems from misunderstanding what is the relevant aspect of the lack of distinction between past and future.}  Hence they alone cannot imply any time-oriented causation.  This fact has been emphasized by Bertrand Russell, who opens his influential 1913 article \emph{On the notion of cause}, claiming that ``\ `cause' is so inextricably bound up with misleading associations as to make its complete extrusion from the  philosophical vocabulary desirable."\cite{Russell1913} 

The idea that causation is nothing other than correlation and that the distinction between `cause' and `effect' is nothing other than the distinction between what comes first and what comes next in time can be traced to Hume, for whom causation is ``an object precedent and contiguous to another, and where all the objects resembling the former are placed in like relations of precedency and contiguity to those objects that resemble the latter", that is, correlations between contiguous events. (Hume is actually subtler in the Treatise: he identifies causation not with the correlation itself, but with the idea in the mind that is determined by noticing these correlations: ``An object precedent and contiguous to another, and so united with it, that the idea of the one determines the mind to form the idea of the other, and the impression of the one to form a more lively idea of the other" \cite{Hume1736}. Even more explicitly in the Enquiry: ``custom ... renders our experience useful to us, and makes us expect, for the future, a similar train of events with those which have appeared in the past."\cite{Hume} I will come back to this in the following.) 

But the usage of the notion of cause, pace Russell, is far from fading, and not just from the philosophical language but not from the scientific language either, and, pace Hume, is not the same as correlation.  In science, as eloquently emphasized for instance by Nancy Cartwright,  not only causal explanations are ubiquitous, but they play an essential role, which is distinct from correlations \cite{Cartwright2007a}. For instance, smoking and entering a hospital are both correlated with death by lung cancer, but it is essential for science to distinguish the fact that smoking causes cancer,  entering the hospital does not.  

However, this does not mean that causation must be a fundamental notion that plays a role in nature at the elementary level (it does not, as Russell \emph{correctly} pointed out), nor that the difference between causation and correlation cannot be understood in terms of something weaker than the existence of irreducible causal laws in our universe (pace Cartwright): it can, as I argue below. Even less (again pace Cartwright), that causation testifies for the disunity of science. What follows, if anything, is an argument for the unity of science.   What all this means is simply that causation is an important concept that we use widely and appropriately, even if it does not make sense at the elementary physical level. Many concepts are like that; for instance: ``cat".  So, what do we mean by causation, given that there is no causation at the elementary physical level? 

Some relevant light on what we actually mean by causation in this sense has been shed by the introduction of causal modelling techniques \cite{ZaltaUriNodelmanColinAllenRLanierAnderson2019}.  These provide concrete working algorithms for discovering relations that are specifically causal, hence making the notion of causation precisely defined.    The way this works can be briefly synthetized as follows.  Assume we have a number of variables $A, B, C, D, ... $ that are partially ordered in time, and we know, --experimentally or  theoretically-- that the values that these variables can take are correlated.  We know these correlations.  Then we say that the variable $B$ is causally related to a variable $C$ that follows it in time if the following happens.  Imagine that we disregard all the correlations between $B$ and \emph{earlier} variables, that is, we allow them to violated, and imagine that we  \emph{intervene} setting the variable of $B$ by hand at one or another value. Then we say that $B$ is causally related to $C$ if the (known) correlations in the future of $B$ imply that  differences in the values of $B$ affect the values of $C$.   

In the smoking example, this signifies that to say that smoking is a cause for cancer, while going to the hospital is not, \emph{means} that if we intervene by preventing people from smoking then
we expect the incidence of lung cancer to decrease, but not so if we prevent people from going to the hospital.  This is the meaning of the statement that smoking causes cancer while going to the hospital does not, even if both are positively correlated with cancer.  

The clarification of the notion of causation brought by the causal modelling techniques is enlightening. Let us analyze it with care.   It is based on two ingredients: the first is the notion of intervention. The second is the idea that intervention affects the future, not the past.  Let us consider the two points in detail. 

\section{Intervention}

In the above analysis, intervention is an interaction between the set of variables considered, $A, B, C, D, ... $ and something else, which upsets the relations between these variables coded in the correlations they have without intervention.   What is this something else?  

In many applications and presumably in the intuition at the basis of the notion of causation, intervention describes a manipulation by a human agent.  I do something and my action causes an effect.  

But the notion of intervention does not require such a strong anthropocentric interpretation.  We equally say that a meteorite falling on the moon causes the formation of a crater, or that intense volcanic activity during a certain geological era is the cause of the presence of a certain chemical in a geological stratum at later times.   In these cases, as in many others, the agent intervening is not human, not biological: it is a meteorite, or volcanic activity.   

What is sufficient to define the notion of an intervention is the fact that we are considering a certain system (the surface of the moon, the surface of the earth, the door opening, Pompei, Kennedy, the body, the pond ...) of which we assume we know --at least to a certain extent-- dynamics and correlations between events, and we are mentally distinguishing this system from something interacting with it (determining the fall of the meteorite, the volcanic activity, the shots by Lee Oswald, something preventing smoking, the fall of the stone...) {whose dynamics we disregard} and which we treat as external agents.  The causal modelling techniques treat the system as a set of correlated variables, but treat the agent as acting arbitrarily, or ``freely". Which is to say: {\em they disregard {its} dynamics}.  This will prove crucial in understanding causation.

\section{Time orientation}

Now, the key assumption in play in  causal modelling is that the intervention of the external agent changes the future, not the past.  This is not implied by the correlations: it is \emph{assumed} in the definition of causation: intervention is assumed to violate \emph{past} correlations, while \emph{future} ones are preserved and determine the effect. 

Why do we assume this, if the elementary laws of nature are time reversal invariant (they do not make any distinction between the past and the future direction of time)?\footnote{For an arbitrary invertible evolution,  causal relations may not be symmetric.  For instance, for $f:(A,B)\to(C=A,D=A+B)$, the variables $A$ and $D$ are causally connected, for but the inverse evolution $f^{-1}:(C,D)\to(A=C,B=D-C)$, they are not.   However this does not happen in classical and quantum mechanics: $f$ is not time-reversal invariant. Thanks to Robert Spekken for making me notice this point.}

The reason is obvious: the laws of nature are time reversal invariant, but in the world around us agency does nevertheless affect the future, not the past.  What does this exactly mean? After all, instead of considering how the future would change if the past was the same and the intervention had not occurred, we could equally consider how the past would change if the future was the same and the intervention had not occurred. That is, we could consider the question of what would happen to the past if we cut ties to future variables and asked how present interventions would affect the past.\footnote{We do use such past counterfactuals:  the light is on, and you tell me that you have just turned it on; if you hadn't (the light being on would have implied that) we left it on earlier, when we last left home. In these cases, we consider different pasts with the same future, determined by the intervention.}  Yet, such reverse logic does not work with causation. Why? 

To see why this is the case, and understand where the time orientation is coming from, consider an example: the history of a pond hit by a stone at a certain location $O$ at time $t=0$. Round concentric waves form around $O$ after the impact: they move outwards and agitate the pond for a while, until their energy dissipates and the pond goes back to a quiet equilibrium state.  Let's treat the fall of the stone as an external intervention, that might or might not have occurred.  Consider the case in which the stone did not fall, and ask what would have happened to the pond, according to the physics we know.   To determine a history, physical laws need the state of the system to be specified at some time; on the basis of this, they determine the history of the system \emph{in both time directions}.  So we have a choice: we can ask what would have happened to the pond if the stone had not fallen 
\begin{enumerate} 
\item[(a)] if the history of the pond had been the same for all times $t<0$, or
\item[(b)] if the history of the pond had been the same for all times $t>0$.
\end{enumerate} 
Physics answers in both cases. In both, the answer is a physical history which is consistent with elementary mechanics.  But the two cases are nevertheless remarkably different in our experience: 
\begin{enumerate} 
\item[(a)] In the first,  nothing remarkable happen for $t>0$: the surface of the pond remains quiet, instead of being excited in concentric waves. 
\item[(b)] In the second, we have to find a past history $h$ of the pond (never hit by anything) such that from time $t=0$ onward there are spherical outgoing waves from a certain location.  This history certainly exists (the water of the pond might have moved in all sort of ways in the past), but it looks strangely implausible to us.  Why so?    Because we know from experience that the water of the pond tends to equilibrate and go back to an equilibrium state.  How come the water of the pond is still agitated in this strange and unnatural manner, if nothing had happened?  
\end{enumerate} 
Thus, among the possible alternatives that we can consider could have happened if the intervention had not occurred, only the ones {\em with the same past} are compatible with the time oriented world in which we happen to live. (That world is ``closer to actuality" \cite{Lewis1973}.) 

The key point is that this argument is not about the elementary mechanical laws of physics: these are compatible with the history $h$.  It is something else that makes $h$ strange.  What is it?
It is thermodynamics: we know that systems equilibrate in time.   Thermodynamics has more ingredients than elementary physics: a macroscopic description of the world and the observation that entropy was low in the past.\footnote{Here by `thermodynamics' I mean also the tendency of systems to equilibrate towards the future, following past lower entropy.  Wayne Myrovold distinguished thermodynamics from equilibration \cite{myrvold2020science}.}

A thermodynamic account of the intervention, in fact, clarifies the source of the asymmetry in time: The pond is in a near equilibrium state a certain temperature $T$.  The stone that hits the pond has a kinetic energy $E$ much larger than the average kinetic energy per degree of freedom of the water molecules, namely $E\gg kT$, where $k$ is the Boltzmann constant.  Hence, the stone is out of equilibrium with the pond. Hence thermodynamics predicts  (probabilistically!) that energy is transferred from the stone to the pond. More precisely, at time $t=0$ energy is transferred to water molecules in the vicinity of $O$.  The transferred energy is free energy for the pond --it is the energy of the outgoing concentric waves-- and thermodynamics tells us that (probabilistically!) this free energy is going to be dissipated into the water, raising its temperature and moving towards energy equipartition.  

This account makes clear why the effect of the intervention is in the future of the intervention and not in its past: because it is in the past that the overall system was away from equilibrium: the stone had more energy than what equipartition of energy indicates.  Because of this past low entropy, the thermodynamical evolution of the system is time oriented, as for all thermodynamical systems with past low entropy.

Thus, the effect (the waves) \emph{follows} the cause (the impact of the stone) instead of preceding it because of the time orientation given by the existence of a thermodynamic disequilibrium \emph{in the past}. 

The waves, namely the effect, follow the fall of the stone, namely the cause, because it is a step in the thermodynamic equilibration of an initial unbalance.  

\section{The thermodynamic arrow is essential}

The structure of the relation between what we call a cause and what we call its effect illustrated above with the example of the stone in the pond is in fact generic: its time orientation is given by the thermodynamic arrow of time, namely the actual fact that entropy was lower in the past.  Causation is therefore a macroscopic thermodynamic phenomenon where the total entropy is raised by an intervention, and the effect is the trace left on the system by the intervention. 

This thermodynamic structure is the same as that characterizing \emph{traces} in general and \emph{memory} in particular \cite{Rovelli2020}.  A trace or a memory are indeed effects of the event they record, which is their cause.   This is also the general structure of the phenomenon we call ``agency" \cite{Rovelli2020a}.  More on this later. 

Can this be proven in general? Yes and the proof is surprisingly simple.  Precisely because causation does capture a time oriented relation, it can only depend on the single source of time-orientation that is compatible with physics: the thermodynamic arrow of time.  Namely the fact that we live in a universe that we describe macroscopically and which has a consistent temporal orientation of its entropy gradients.\footnote{Attempts to trace the arrow of time to something else (such as \cite{Gold2005}) are either unconvincing, or can themselves be traced to the statistical irreversibility connected to low entropy in the past. On perspectival grounding for irreversibility \cite{Price2023}, see the discussion below, in Section \ref{agent}.} 

To further clarify this fact, consider the two extreme situations where there is no entropy gradient.  The first is when the overall system is at or near thermal equilibrium.  In this case, the energy of the stone falling into the pond must be of the order $kT$.  If so, its effect on the water molecules is indistinguishable from the generic thermal agitation: the fall has no detectable effect.    There is no time orientation and no sense of causation. 

The other case is the case of purely mechanical interactions, without any coarse-grained description.  In this context, there is no notion of entropy, no notion of more or less probable macroscopic states, and therefore no sense in which considering histories with the same future can be less plausible than considering histories with the same past.  There is no intrinsic distinction between past and future and therefore, again, no possible intrinsic time oriented causation.   If two stones with velocities $v_1$ and $v_2$ collide and after the collision have velocities $w_1$ and $w_2$, then there is nothing in the phenomenon itself that fixes a preferred time orientation. We can equally say that without the collision the velocities would have been always  $v_1$ and $v_2$ or always $w_1$ and $w_2$.  If we say that the \emph{later} velocities are caused by the collision, we are truly in the situation described by Hume: we notice a correlation and the term we call ``cause" is only characterized by the fact that it happens earlier: causation in this sense is reduced to (symmetric) correlation.

Is that all? No. I believe we are are still missing \emph{the} crucial ingredient of this story. 

\section{The agent's time arrows}\label{agent}

If the above is physically correct, why is it of any relevance?  As defined above, causation is an intricate and baroque thermodynamic phenomenon describing certain peculiar statistical patterns in the interaction between systems. Why is then causation so important in our making sense of the world?  

As Huw Price puts it (tracing the idea to Ramsey \cite{Ramsey1978}): the interesting question isn't what causation is, but how we come to think and talk in causal terms \cite{Price2023}. In other words, we do not understand causation by asking which complicate patterns in the tapestry of nature we happen to call causation; we understand causation by understanding why those patterns are relevant at all; which is to say, by understand what is causal thinking.

In the previous pages, there were numerous hints about the answer: (i) the intuition about the notion of intervention comes from \emph{our own} capacity of intervening and manipulating systems; (ii) we are interested in the fact that it is smoking to cause cancer because because \emph{we can actually intervene on this}; (iii) as Hume points out, causation is not really something happening in the phenomena: it is \emph{the idea in the mind that we gather from that which is useful to us}.  All this points to the fact that we use causation as a predictive tool for possible futures, where \emph{we ourselves are the agents} that can intervene.   

\emph{We} are subjects of knowledge and actors in the world that have a direct involvement in the causation game and a direct interest in causal relations.  After all, our brain is essentially a machine that analyses the different possible futures that would follow {\em if this or that course of action is taken}. The main business our brain is involved in  is not simply to predict the future given the past \cite{Buonomano2017}, but to predict what would happen under different choices of behaviour, namely to predict what would the effects of different interventions be. 

A (neo-)pragmatist perspective is therefore a natural context to understand why causation is key to our understanding of the world.  That is, pragmatism is particularly convenient for explaining the relevance of causation, and what is causation in the perspective of the subject, as clarified by Huw Price \cite{Price2007,Price2023}.  Citing Ramsey again: ``from the situation when we are deliberating seems to me to arise the general difference of cause and effect."  \cite{Ramsey1978}, 

From such neo-pragamatist perspective, however, we have to investigate the subject that utilizes causation in this manner, and which determines its relevance. The subject is of course itself a natural being subjected to the laws of physics.  Its peculiar behaviour must therefore be grounded into physics. In particular, its own time orientation cannot but itself be grounded in physics.  Let's analyze this  fact in detail.   
\begin{itemize}
\addtolength{\itemsep}{-2mm}
\item[(a)] The subject knows the past, not the future.
\item[(b)] The subject can choose which actions to do. 
\item[(c)] Its choice affects the future, not the past. 
\end{itemize}
The first is the epistemic arrow of time (we know the past better than the future). The second is the vivid intuition of our freedom in choosing, which sometimes goes under the name of free will. The third is the agential arrow of time (we can affect the future, not the past).   Crucially, the time orientation of these aspects of the subject's phenomenology are themselves effects of the thermodynamic arrow of time.    

The epistemic arrow of time is a consequence of the fact that an entropy gradient plus some additional simple conditions largely realized in our universe (long thermalization times and systems' separation) are sufficient for the formation of abundant traces of the past (that have no time reversed equivalent.)  The past is fixed because in the present there are traces about it, and these are there because of entropy was low in the past. This was anticipated in \cite{Reichenbach1956} and is discussed in detail in \cite{Rovelli2020}. 

The possibility for the agent of determining different macroscopic futures is also permitted by an entropy gradient.  There is a simple way to prove this \cite{Rovelli1956}. A choice is a (physical) process which macroscopically can evolve into two different futures (with the same past). This is not in contradiction with determinism \cite{Spinoza1677b} because it is a macroscopic description: different micro-histories that in the past were part of the same macro-state can evolve into different macro-states \cite{Rovelli2020a,Loewer2020}.   However, if we look at the macro-physics only, there is a gain of information at the choice: the information about which choice was actualized.  This new macroscopic information cannot come for free: it must be paid for in entropy increase, that is, dissipation (which loses macroscopic information). {\em  Hence any agent that chooses necessarily dissipates free energy into heat.}  Hence agency is a macroscopic thermodynamic phenomenon that gets its time orientation from the entropy gradient.  Only in this macroscopic context the notion of causation makes sense, the context where thermodynamics is relevant and time orientation makes sense.  

Intervention was defined by assuming the intervening agent to be `free': this is possible precisely because it amounts to disregarding some degrees of freedom and their dynamics. Our own agency is in part grounded in the same ignorance of  degrees of freedom and their dynamics \cite{Spinoza1677b}. More specifically, however, biological agents determine their own behaviour in part by calculating future outcomes of their possible alternative.  The embedded (physical!) information they utilize is itself part of what determines the future and as such it interferes (in the sense of \cite{Ismael2022}) with what is going to happen. For the agent's (embedded, physical) representation of the world, the future is therefore necessarily `open': it depends on its choices \cite{Ismael2022}.  

The opposite perspective, which takes into account how came we do so, is probably even more clarifying: the entropy gradient of the world creates opportunities for the use of information to guide behaviour.  So, we  have evolved to pick out from the environment macroscopic variables to interact with, suitable for us to act upon the world in ways that we can (partially) causally control to our own advantage \cite{Ismael2022a}.

In sum, causality is indeed better understood --I believe-- not as a simple physical issue, and even less as a metaphysical one; but rather in terms of the perspective and interest of deliberating human agents. But the question of what is it in the natural world that enables us humans to use it is a well posed issue, and a scientific one. The answer is the (approximate, contingent) entropy gradient.    From the physical perspective, all such features of the subject that motivates its interest in causation are therefore themselves rooted in the thermodynamic arrow of time.  

In closing, let's get back to the situation where we interpret correlation as causation in settings where there is no dissipation, namely no thermodynamic time orientation in the phenomena themselves. For instance in the case of two particles colliding, for which the velocities were $v_1$ and $v_2$ before the collision and are $w_1$ and $w_2$ after the collision. In these case we \emph{still} say that the collision \emph{caused} the velocities to become $w_1$ and $w_2$. Not (as argued above) because of an intrinsic time orientation in the particle's phenomenology, but because of the time orientation of us subjects, itself rooted in the entropy gradient. (On this, see also Section 5 of  \cite{Price2007}.)  Causation is largely pertaining to the perspective of a deliberating agents. \\

\section{Conclusion}

According to current physics, the basic laws of our universe are time reversal invariant: they do not determine a preferred direction of time. They express correlations between events, which are symmetric in time.  There is no notion of causation at this level. This came as a historical surprise. 

On the other hand,  the state in which the universe happens to be is not symmetric under time reversal.  We access it only via a relative small number of degrees of freedom, and the entropy that such coarse graining defines happened to be far lower in the direction of time that we call the past.  There is a commonly oriented entropy gradient in all phenomena we witness, as well as in our own behaviour. 

A consequence of this state of affairs is that the present contains abundant traces of the past that have no time reversed equivalent.  Another consequence is that  biological critters like us collect information about the past and utilize it to determine their behaviour, affecting the future by intervening on physical processes.  We say that any intervention affects the future and not the past in the macroscopic world because we are only concerned with macroscopic histories consistent with the entropy gradient we witness around us. 

As Huw Price points out (attributing the insights to Ramsey) ``Pace Cartwright, we get to this distinction [the distinction between causation and correlation] not by adding causal laws to bare associations in the ontological realm, but by taking something away from the import of those associations in the epistemic realm. Ramsey's great insight is to see how deliberation does the subtraction for us"  \cite{Price2023}. I think this is right. But things are subtle and I would add that our capacity of deliberation may well lead us to conceive and value causation, but once this is place, we do not need deliberation to talk about causation. There are ways weaker than deliberation of ``taking something away from the import of correlations in the epistemic realm": simply \emph{treating} a system as an agent by ignoring its own history, as we do for a meteorite that falls on the moon. We do not know when it falls, but we do not attribute deliberation to it.  More importantly, the feature of nature (no causation here) that underpin the fact that (what we call) the effect of the meteorite as well as (what we call) our own deliberation, act on the future, hence enabling us to talk in causal terms, is the ubiquitous thermodynamical arrow of time.   

Hence the arrow of causation is ultimately rooted into the thermodynamic arrow via two different paths. Because in the macrophysics of our actual world, intervention does in fact affect the future, in the sense that considering it affecting the past generates thermodynamic implausible hence irrelevant histories. And because causation is the concern of the biological systems we are, time-oriented by the biosphere's entropy gradient.  For this second reason, causation is something \emph{we} read in the worlds.  For the first is something we can read \emph{in the world}.

Our own thinking is a dissipative process,  time-oriented by the entropy gradient. So it is hard for our intuition to accept the fact that time orientation and hence the arrow of causation, are only thermodynamic, that is  statistic, approximate, perspectival, phenomena.  Our entrenched intuition notwithstanding, it is so. \\ 

\centerline{***}

Sincere thanks to Jenann Ismael and to Emily Adlam for sharing ideas that had a strong influence in this work and for a carful reading of the draft.  And to Huw Price for useful exchanges and sharing his manuscript. 


\begin{thebibliography}{10}

\bibitem{Lewis1979}
D.~K. Lewis, ``{Counterfactual Dependence and Time's Arrow},'' {\em Nous}
  {\bfseries 13} (1979) 455--476.
  \url{https://joelvelasco.net/teaching/5330(fall2013)/lewis79-CFTimesArrow.pdf}.

\bibitem{Dowe1975}
P.~Dowe, ``{Process Causality and Asymmetry},'' {\em Erkenntnis} {\bfseries 37}
  (1975) 179--196.
  \url{https://www.jstor.org/stable/20012433#metadata_info_tab_contents}.

\bibitem{Ismael2022a}
J.~Ismael, ``{It's not what you look at, it's what you see},'' in {\em Causal
  Perspectivalism Conference – June 13-14 2022}.
\newblock 2022.
\newblock
  \url{https://framephys.org/causal-perspectivalism-conference-june-2022/}.

\bibitem{Carroll}
S.~Carroll, ``{The Arrow of Time in Causal Networks}.''
\newblock \url{https://www.youtube.com/watch?v=6slug9rjaIQ}.

\bibitem{Falcon2011}
A.~Falcon, ``{Aristotle on Causality},'' {\em Stanford Encyclopedia of
  Philosophy} (2011) 1--14.
  \url{http://plato.stanford.edu/archives/fall2011/entries/aristotle-causality/}.

\bibitem{Dhammajoti2009}
B.~Dhammajoti, {\em {Sarvāstivāda Abhidharma.}}
\newblock Centre of Buddhist Studies, The University of Hong Kong., 2009.

\bibitem{Adlam2022}
E.~Adlam, ``{Is There Causation in Fundamental Physics? New Insights from
  Process Matrices and Quantum Causal Modelling},''
  \href{http://arxiv.org/abs/2208.02721}{{\ttfamily arXiv:2208.02721}}.
  \url{https://arxiv.org/abs/2208.02721v1 http://arxiv.org/abs/2208.02721}.

\bibitem{Russell1913}
B.~Russell, ``{On the Notion of Cause , 1912 - 1913},'' {\em Proceedings of the
  Aristotelian Society} {\bfseries 13} (1913) 1--26.

\bibitem{Hume1736}
D.~Hume, {\em {Treatise of Human Nature}}.
\newblock 1736.
\newblock \url{https://www.gutenberg.org/files/4705/4705-h/4705-h.htm}.

\bibitem{Hume}
D.~Hume, {\em {Enquiries concerning Human Understanding and concerning the
  Principles of Morals}}.
\newblock 1777.
\newblock \url{https://www.gutenberg.org/ebooks/9662}.

\bibitem{Cartwright2007a}
N.~Cartwright, \href{http://dx.doi.org/10.1017/CBO9780511618758}{{\em {Hunting
  causes and using them: Approaches in philosophy and economics}}}.
\newblock Cambridge University Press, Jan, 2007.

\bibitem{ZaltaUriNodelmanColinAllenRLanierAnderson2019}
E.~N. {Zalta Uri Nodelman Colin Allen R Lanier Anderson}, ``{Stanford
  Encyclopedia of Philosophy Causal Models},'' 2019.
\newblock \url{https://plato.stanford.edu/entries/causal-models/}.

\bibitem{Lewis1973}
D.~Lewis, ``{Causation},''  The
  Journal of Philosophy {\bfseries 70} no.~17, (Oct, 1973) 556.
  \url{https://philpapers.org/rec/LEWC}.

\bibitem{myrvold2020science}
W.~C. Myrvold, ``{The Science of $\Theta\Delta$},''
 Foundations of
  Physics {50} (2020) 1219--1251,
  \href{http://arxiv.org/abs/2007.11729}{{\ttfamily arXiv:2007.11729}}.

\bibitem{Rovelli2020}
C.~Rovelli, ``{Memory and entropy},'' {\em Entropy 2022} {\bfseries 24} (2022)
  1022, \href{http://arxiv.org/abs/2003.06687}{{\ttfamily arXiv:2003.06687}}.

\bibitem{Rovelli2020a}
C.~Rovelli, ``{Agency in Physics},'' in {\em Experience, abstraction and the
  scientific image of the world.}
\newblock Franco Angeli editore, 2021.
\newblock \href{http://arxiv.org/abs/2007.05300}{{\ttfamily arXiv:2007.05300}}.

\bibitem{Gold2005}
T.~Gold, ``{The Arrow of Time},''
 American Journal of Physics
  {\bfseries 30} no.~6, (Jul, 2005) 403.
  \url{https://aapt.scitation.org/doi/abs/10.1119/1.1942052}.

\bibitem{Price2023}
H.~Price, ``{Time for Pragmatism},'' in {\em Neo-pragmatism}, J.~Gert, ed.
\newblock Oxford University Press, 2023.

\bibitem{Ramsey1978}
F.~Ramsey, ``{General Propositions and Causality},'' in {\em Foundations:
  Essays in Philosophy, Logic, Mathematicvs and Economics}, D.~Mellor, ed.,
  pp.~133--151.
\newblock Routledge anf Kegan, London, 1978.

\bibitem{Buonomano2017}
D.~Buonomano, {\em {Your Brain Is A Time Machine: The Neuroscience and Physics
  of Time}}.
\newblock W W Norton and Co Inc, 2017.

\bibitem{Price2007}
H.~Price, ``{Causal perspectivalism},'' in {\em Causation, Physics, and the
  Constitution of Reality: Russell's Republic Revisited}, pp.~250--292.
\newblock 2007.
\newblock \url{https://philpapers.org/rec/PRICP}.

\bibitem{Reichenbach1956}
H.~Reichenbach, {\em {The Direction of Time}}.
\newblock Dover, New York, 1956.

\bibitem{Rovelli1956}
C.~Rovelli, ``{Back to Reichenbach},''.
  \url{http://philsci-archive.pitt.edu/20148/}.

\bibitem{Spinoza1677b}
B.~Spinoza, {\em {Ethics, Third Part, Proposition II and Scolio}}.
\newblock 1677.
\newblock \url{http://en.wikisource.org/wiki/Ethics_(Spinoza)/Part_3}.

\bibitem{Loewer2020}
B.~Loewer, ``{The Consequence Argument Meets the Mentaculus},''.
  \url{http://philsci-archive.pitt.edu/17328/}.

\bibitem{Ismael2022}
J.~Ismael, ``{The Open Universe: Totality, Self-reference and Time},'' {\em
  Australasian Philosophical Review} (2022) to appear.

\end{thebibliography}

\providecommand{\href}[2]{#2}\begingroup\raggedright\endgroup

\end{document}